\documentclass[twocolumn,secnumarabic,amssymb, nobibnotes, aps, pra,showpacs,preprintnumbers]{revtex4-1}
\usepackage{amssymb}
\usepackage{array}
\usepackage{graphicx}
\usepackage{amsmath}
\usepackage{amsthm}
\usepackage{amsopn}
\def\uno{\mbox{1 \kern-.59em {\rm l}}}

\def\beq{\begin{equation}}
\def\eeq{\end{equation}}
\def\bea{\begin{eqnarray}}
\def\eea{\end{eqnarray}}

\begin{document}

\title{Constraining f(T) gravity by dynamical system analysis}%

\author{Behrouz Mirza}
\email{b.mirza@cc.iut.ac.ir}
\author{Fatemeh Oboudiat}
\email{f.oboudiat@ph.iut.ac.ir}
\affiliation{Department of Physics,
Isfahan University of Technology, Isfahan 84156-83111, Iran}

\begin{abstract}
We investigate the cosmological solutions of the $f(T)$ gravity theory using the method of dynamical systems. For this purpose a general form of the $f(T)$ function is considered and three conditions are defined that they have to satisfy in order to describe the standard cosmological history. We examine five specific models of $f(T)$ gravity and obtained the valid range of their parameters.

\end{abstract}


\maketitle

\newpage
\section{Introduction}
Recent cosmological observations indicate that our universe has an accelerated expansion \cite{astro}. This is not  predicted by General Relativity (GR) using ordinary baryonic matter. Either of two approaches might be adopted to describe the expansion of the universe: 1) Introducing a new kind of matter, called dark energy, into the GR theory  which either raises problems such as future singularities or violates certain energy conditions; and 2) Modifying GR to some new theory of gravity like $f(R)$ \cite{fr} or other forms of higher derivative theories \cite{high}.
\\ An equivalent formulation of GR is teleparallel gravity (TG) in which torsion, rather than  curvature, is responsible for gravitational interaction \cite{teleparallel}. Originally due to Einstein, TG which is basically different from GR was meant  to unify electromagnetism and GR \cite{TEGR}. TG is considered a gauge theory \cite{gauge} while GR is described as a geometric one; however, the same equations of motion apply to both; hence, the designation Teleparallel Equivalent of General Relativity (TEGR).
\\ One straightforward modification of TEGR is $f(T)$ gravity in which the  torsion scalar $T$ is replaced with $f(T)$ in the action \cite{ft} similar to $f(R)$ gravity in which the Ricci scalar is replaced with $f(R)$ in the action (for a review see \cite{revft}). Compared to its GR equivalent $f(R)$ theory, its equations of motion are of second order in contrast to the fourth order equations of motion in $f(R)$ theory but there is no local Lorentz invariance so it is not possible to fix some of the vierbeins by gauge symmetry. The theory can explain the present cosmic acceleration \cite{acceleration} and solve the problem of inflation with no inflaton \cite{inflation}. An account of the cosmological evolution of the theory may be found in \cite{cosmology} while  \cite{scalar} may be consulted for the advantages of adding a scalar field to the theory. Observational data have been used to constrain the model parameters in Ref. \cite{observ1} and the Noether symmetry is examined in \cite{Neother}. Finally, the possibility of wormhole is investigated in \cite{wormhole} and the phantom divide crossing is studied in \cite{phantomcross}.
\\The  autonomous dynamical system is a tool for investigating the fixed points and singularities of a theory \cite{fix}. This method is examined for $f(T)$ theory in \cite{fix ft} where a special form of $f(T)$ is selected. A general form of $f(T)$ function is used in \cite{observation}, but the calculations and conclusions therein are not correct. In this paper, we study $f(T)$ gravity using the dynamical system method in by the general form of $f(T)$ function. We find three conditions to constrain the functional form of $f(T)$.
\\ The paper is organized as follows. In Section II, we briefly explain the fundamentals of TG and the action of the $f(T)$ theory. Section III examines the method of autonomous dynamical system for a general form of $f(T)$ theory. Some special cases are presented in Section IV. Summary and conclusions are finally presented in Section V.

\section{Equations of motion}
In this section, we briefly review the equations of motion of $f(T)$ gravity. The dynamical variables of TG and $f(T)$ gravity are vierbein fields $e_{A}$, $A=  0,1,2,3$, which form an orthogonal basis for the tangent space at each point $x_{\mu}$ of the manifold so that $e_{A}.e_{B}=  \eta_{AB}$, where $\eta_{AB}=  diag(1,-1,-1,-1)$. The vector $e_{A}$ can be described by its components on a coordinate basis, $e_{A}=  e_{A}^{\mu}\partial_{\mu}$, where the Latin indices refer to the tangent space and the Greek ones to the coordinate space on the manifold. The metric is defined as  $g_{\mu\nu}=  \eta_{AB}e^{A}_{\mu}(x)e^{B}_{\nu}(x)$.
\\Instead of the torsion-less Levi-Civita connection in the General Relativity, we use a curvatureless Weitzenb\"{o}ck connection \cite{weits}:
\bea
\Gamma_{\mu\nu}^{\lambda}\equiv  e_{A}^{\lambda}\partial_{\nu}e_{\mu}^{A}.
\eea
The torsion tensor from this connection reads:
\bea
T_{\mu\nu}^{\lambda}=  \Gamma_{\nu\mu}^{\lambda}-\Gamma_{\mu\nu}^{\lambda}=  e_{A}^{\lambda}\left(\partial_{\mu}e_{\nu}^{A}-\partial_{\nu}e_{\mu}^{A}\right)\label{tor}.
\eea
Using torsion tensor (\ref{tor}), we can construct the contorsion tensor and the $S$ matrix as follows:
\bea
K^{\mu\nu}_{\:\:\:\:\rho}&=  &-\frac{1}{2}\Big(T^{\mu\nu}_{\:\:\:\:\rho}-T^{\nu\mu}_{\:\:\:\:\rho}-T_{\rho}^{\:\:\:\:\mu\nu}\Big),\\
S_\rho^{\:\:\:\mu\nu}&=  &\frac{1}{2}\Big(K^{\mu\nu}_{\:\:\:\:\rho}+\delta^\mu_\rho \:T^{\alpha\nu}_{\:\:\:\:\alpha}-\delta^\nu_\rho\:
T^{\alpha\mu}_{\:\:\:\:\alpha}\Big),
\eea
both of which are antisymmetric tensors. Using these quantities, the torsion scalar $T$ can be defined as follows:
\bea
T\equiv  S_\rho^{\:\:\:\mu\nu}\:T^\rho_{\:\:\:\mu\nu} \label{T},
\eea
which is the Lagrangian density of the so-called TG. \\
The idea $f(T)$ gravity is to extend $T$ to $f(T)$, similar to the generalization of $R$ to $f(R)$ in Einstein-Hilbert action. So, the action of the $f(T)$ gravity reads:
\bea
I=  \frac{1}{2k}\int d^{4}x e\left(T+f(T)+\mathcal{L}_{m}+\mathcal{L}_{r}\right),\label{action}
\eea
where, $k=  8{\pi}G$, $G$ is the Newtonian constant, $e=  det(e_{\mu}^{A})=  \sqrt{-g}$, and $\mathcal{L}_{m}$ and $\mathcal{L}_{r}$ are the matter and radiation lagrangian densities. Varying the action with respect to the vierbein, we have:
\bea
&&e^{-1}\partial_{\mu}(ee_A^{\rho}S_{\rho}{}^{\mu\nu})[1+f'({T})]-e_{A}^{\lambda}T^{\rho}{}_{\mu\lambda}S_{\rho}{}^{\nu\mu}[1+f'({T})]\nonumber\\&& +e_A^{\rho}S_{\rho}{}^{\mu\nu}\partial_{\mu}({T})f''({T})-\frac{1}{4}e_{A}^{\nu}[T+f({T})]=  \frac{k}{2}e_{A}^{\rho}T_{\rho}{}^{\nu},\quad
\label{eom}
\eea
whose prime means differentiation with respect to the label and $T_{\rho}{}^{\nu}$ is the energy momentum tensor.
\\We choose the vierbein:
\bea
e_{\mu}^A=  {\rm diag}(1,a,a,a),\label{vierbein}
\eea
which leads to the flat FRW universe by the metric:
\bea
ds^2=   dt^2-a^2(t)dx^{2}.
\eea
Putting the veirbein (\ref{vierbein}) in (\ref{T}), the torsion scalar will take the following form:
\bea
T=  -6H^2,
\eea
where, $H=  \frac{\dot{a}}{a}$ is the Hubble parameter. Assuming that the matter content of the universe is a perfect fluid with the energy momentum tensor $T_{\mu\nu}=   p_mg_{\mu\nu} - (\rho_m +p_m)u_{\mu}u_{\nu}$, we can find the Friedmann-like equations from (\ref{eom}) as follows:
\bea
&&H^2=   \frac{k}{3}(\rho_m+\rho_r) -\frac{f({T})}{6}-2f'({T})H^2,\label{frid1}\\
&&\dot{H}=  \frac{-\frac{k}{2}\left(\rho_m+p_m+\rho_r+p_r\right)}{1+f'(T)-12H^2f''(T)},\label{frid2}
\eea
where, $\rho_m$ and $p_m$ are the density and pressure of the matter while $\rho_r$ and $p_r$ are the density and pressure of the radiation. The superscript dot denotes the derivative with respect to the cosmic time $t$.

\section{Autonomous dynamical system in its general case}
We consider the $f(T)$ gravity theory in the flat isotropic and homogeneous FRW space-time. The matter content of the universe is chosen to be dust (with an energy density of $\rho_m$ and zero pressure). Radiation parameters, $\rho_r$ and $p_r$, are related to each other via the equation of state $p_r=  \omega_r\rho_r$ whose constant is $\omega_r=  \frac{1}{3}$. No interaction is assumed to occur between radiation and matter; so, the continuity equations for matter and radiation will read as follows:
\bea
&&\dot{\rho}_m+3H\rho_m=  0,\nonumber\\
&&\dot{\rho}_r+4H\rho_r=  0.\label{contin}
\eea
Assuming that dust and radiation are present in the flat FRW universe, the Friedman equations (\ref{frid1}) and (\ref{frid2}) will change to (\ref{fr1}) and (\ref{fr2}) below:
\bea
&&H^2=   \frac{k}{3}(\rho_m+\rho_r) -\frac{f({T})}{6}-2f'({T})H^2,\label{fr1}\\
&&\dot{H}\left(1+f'(T)-12H^2f''(T)\right)=  -\frac{k}{2}\rho_m-\frac{2}{3}k\rho_r .\label{fr2}
\eea
We will consider the general case of the theory with no constraint on the $f(T)$ function in the present section and some useful special cases in the subsections to follow.\\
The method of dynamical systems is one used for investigating the whole dynamics of a system near extremum points of the theory called, fixed points. In this method, the qualitative behavior of the system is pictured in the phase space by some trajectories which depend on initial values. Based on early and late time behaviors of the system as well as radiation and matter solutions, one can choose some of the trajectories and rule out the inconsistent ones. It is appropriate at this juncture to introduce some dimensionless independent variables in order to simplify the calculations:
\bea
\Omega_m&=  &\frac{k\rho_m}{3H^2},\label{omegam}\\
\Omega_r&=  &\frac{k\rho_r}{3H^2},\label{omegar}\\
x&=  &\frac{f(T)}{T},\label{x}\\
y&=  &-2f'(T)\label{y},\\
z&=  &2Tf''(T)\label{z}.
\eea
It is clear from Equation (\ref{fr1}) that the effective density of dark energy is equal to
\bea
\rho_D=  \frac{3}{k}\left(-\frac{f(T)}{6}-2f'(T)H^2\right).
\eea
So, we can define $\Omega_D$ as follows:
\bea
\Omega_D= \frac{k\rho_D}{3H^2}=  x+y\label{Omegad}.
\eea

\begin{table*}
\begin{center}
\caption{\bf The fixed points and physical parameters of the system (\ref{ode}).}
\vspace{5mm}
\begin{tabular}{|c|c|c|c|c|c|c|c|c|c|}
  \hline

  {\bf fixed point}  &  $x$  & $\Omega_{r}$&$\Omega_{m}$ &$\Omega_D$&$q$  &$\lambda_{1}$ & $\lambda_{2}$ &  $y'(x_0)<-2$  &  $y'(x_0)>-2$ \\\hline\hline

$P_{1}$ & $x_0$ & $1+x_0$ &$0$ &  $-x_0$&$1$ & $1$&$2\left(y'(x_0)+2\right)$&saddle&unstable  \\\hline

$P_{2}$ & $x_0$ & $0$ & $1+x_0$&$-x_0$&$\frac{1}{2}$& $-1$&$\frac{3}{2}\left(y'(x_0)+2\right)$&stable&saddle    \\\hline

$P_{3}$ & $x_1$ & $0$ & $0$& $1$&$-1$ & $-3$&$-4$&stable&stable     \\\hline

$P_{4}$ & $x_2$ & $0$ &$1-x_2-y(x_2)$& $x_2+y(x_2)$ &$-1$ & $-4$&$\lambda (x_2)$&---&---  \\\hline

\end{tabular}
\\The point $x_0$ is the solution of $y(x)+2x=  0$, the point $x_1$ is the solution of $y(x)+x=  1$, and the point $x_2$ is the value in which $1-\frac{1}{2}y+y'\left(x+\frac{1}{2}y\right)$ diverges. The fixed points $P_2$ for $y'(x_0)<-2$ and $P_3$ are stable. $P_3$ is the solution of late time accelerating phase of the universe.

\label{general}
\end{center}
\end{table*}

We should make sure that the variables (\ref{omegam}) to (\ref{z}) are independent; if not, an independent set should be chosen from among them. It is obvious from (\ref{x}) that $x=  x(T)$ and, reversely, $T=  T(x)$. Thus, $y(T)$ and $z(T)$ are not independent of $x$ and we have $y(x)$ and $z(x)$. So, from among the three variables $x$, $y$, and $z$, we choose just $x$ as an independent variable. By differentiating $y$ with respect to $x$, we can find an explicit relation between $z$ and $y$ as in the following:
\bea
y'(x)=  \frac{dy}{dx}=  \frac{dy/dT}{dx/dT}=  \frac{-2f''(T)}{\frac{f'(T)}{T}-\frac{f(T)}{T^2}}=  \frac{z(x)}{x+\frac{1}{2}y(x)},\label{y'}\qquad
\eea
where, $y'(x)$ means differentiating $y$ with respect to $x$, i.e. $\frac{dy}{dx}$, not to be confused with $\frac{dy}{dT}$.\\
For our purposes, we choose $x$, $\Omega_m$, and $\Omega_r$ while we eliminate $y$ and $z$. Using the Friedman Equations (\ref{fr1}) and (\ref{fr2}), the variables (\ref{omegam}) to (\ref{z}), and the relation (\ref{y'}), we can write:
\bea
&&\Omega_m+\Omega_r +x+y=  1,\label{ffr1}\\
&&\frac{\dot{H}}{H^2}\left(1-\frac{1}{2}y+z\right)=  -\frac{3}{2}\Omega_m-2\Omega_r. \label{ffr2}
\eea
Based on Eq. \eqref{ffr1}, we find that there are only two independent variables and that the dynamical system is a two-dimensional one. We choose $\Omega_r$ and $x$ as the independent dynamical variables and write the autonomous dynamical system as in the following:
\bea
\frac{d\Omega_r}{dN}&=  &-2\Omega_r\left(\frac{-\frac{3}{2}-\frac{1}{2}\Omega_r+\frac{3}{2}(x+y)}{1-\frac{1}{2}y+y'\left(x+\frac{1}{2}y\right)}+2\right),\nonumber\\
\frac{dx}{dN}&=  &-\frac{-\frac{3}{2}-\frac{1}{2}\Omega_r+\frac{3}{2}(x+y)}{1-\frac{1}{2}y+y'\left(x+\frac{1}{2}y\right)}(y+2x),\label{ode}
\eea
where, $N=  \ln a$. Moreover, both continuity (\ref{contin}) and Friedman Equations (\ref{ffr1}) and (\ref{ffr2}) are used. In Ref. \cite{observation}, $x$, $y$, and $\Omega_r$ are chosen as independent variables. As already explained, $y$ cannot be assumed to be independent from $x$ and the autonomous set consists of $x$ and $\Omega_r$. In order to study the autonomous system around the equilibrium points, the right hand side of the system (\ref{ode}) should be continuous and differentiable. Hence, the  following constraint applies to all the solutions of the system:
\bea
&&1-\frac{1}{2}y(x)+y'(x)\left(x+\frac{1}{2}y(x)\right)=  1+f'(T)+2Tf''(T)\nonumber\\&&\neq 0.\label{exception}
\eea
The left hand side of Equation (\ref{exception}) is the denominator of the dynamical Equations (\ref{ode}). The special case $ 1+f'(T)+2Tf''(T)= 0 $ has a simple solution that will be studied in the next section.\\
Extremum points of the system, called fixed or critical points, occur if $\frac{dx}{dN}=  \frac{d\Omega_r}{dN}=  0$. We find four categories of fixed points as presented in Table \ref{general}.
An accelerated universe requires minus values to be found for the deceleration parameter represented by $q$ in Table \ref{general} for the present phase of the universe. This parameter is defined as follows:
\bea
q=-\frac{a\ddot{a}}{\dot{a}^2}=  -1-\frac{\dot{H}}{H^2}=  -1+\frac{\frac{3}{2}+\frac{1}{2}\Omega_r-\frac{3}{2}(x+y)}{1-\frac{1}{2}y+y'\left(x+\frac{1}{2}y\right)}.\qquad
\eea
It is possible to investigate the stability of the solutions through the study of the eigenvalues of the first order perturbation matrix (Jacobian matrix) near the critical points. These eigenvalues are represented by $\lambda_i$ in Table \ref{general}. The stability of the regimes presented in this Table depends on the values of $y'(x_0)$, where $x_0$ is the solution of  $y(x)+2x=  0$. Depending on the eigenvalues of the Jacobian matrix, we may have different kinds of stability states. The stable, unstable, and saddle states are possible for real values of $\lambda_{i}$. Positive eigenvalues yield an unstable fixed point; negative ones, a stable one; while mixed positive and negative values yield a saddle fixed point. When some or all of the eigenvalues have a zero real part, the fixed point is called nonhyperbolic. Novel theories will be required to determine the stability of nonhyperbolic fixed points as the linear stability theory fails to do so.\\
We are living in a dark-energy-dominated accelerated expanding universe. Going back in time, the universe used to be denser than it is today when matter dominated the universe. Since radiation decays faster than matter in an expanding universe (based on the solution of Equations (\ref{contin})), radiation must have preceded matter in dominating the world. Prior to the radiation-dominated world, an accelerated expansion, called inflation, is believed to have existed, which connected the beginning of the universe Big Bang to the radiation era after inflation. Thus, the universe experienced two eras of acceleration; early time acceleration due to inflation and late time acceleration due to dark energy. Hence, any proposed cosmological model should contain at least part of the cosmological model below \cite{fix}:\\
inflation$\rightarrow$radiation$\rightarrow$matter$\rightarrow$dark energy.\\
To obtain the cosmological model above, inflation should be an unstable point for the universe to have inflation exit while radiation and matter points should be saddle ones in order for it to have long enough radiation and matter eras. Finally, dark energy era should be a stable point to have an expanding accelerated phase at the end.\\
Let us now turn to the theory outlined here and see if it is a viable cosmological model. There are four categories of fixed points in Table \ref{general}. The first category of fixed points (which are represented by $P_1$ in the Table) occur with the solutions of $y(x)+2x=  0$. We call the roots of this equation $x_0$. Depending on the functionality of $y(x)$, it is possible to have zero, one, or more fixed points. For all these points, $\Omega_r=1+x_0 $ and $\Omega_m=0  $. The point changes to a radiation-dominated fixed point for small values of $x_0$ because $\Omega_m=0$ and $\Omega_D=  x+y(x)=0$. Unstable solutions will be obtained in the domain $y'(x_0)>-2$ because both the eigenvalues of the Jacobian matrix (i.e., $\lambda_1$ and $\lambda_2$) are positive. In the same way, saddle solutions are obtained in the domain $y'(x_0)<-2$ because one eigenvalue is positive and the other is negative. Finally, nonhyperbolic solutions are obtained for $y'(x_0)=-2$. For $x_0=-1$, $P_1$ changes to a dark energy-dominated critical point. Since there is no stable solution or accelerating phase, ($q>0$), it fails to describe the late time dark energy-dominated phase of the universe.\\
The second  category of fixed points ($P_2$ in the Table) are again the solutions of the equation $y(x)+2x=  0$ with $\Omega_r=0$ and $\Omega_m=1+x_0$. For small values of $x_0$, $\Omega_m$ tends to one, $\Omega_D$ equals zero, and the point will be a matter-dominated fixed point which will be a stable one for $y'(x_0)<-2$, a saddle one for $y'(x_0)>-2$, and a nonhyperbolic one for $y'(x_0)=-2$. Similar to $P_1$, $P_2$ changes to a dark energy-dominated fixed point for $x_0=-1$, but it is stable for $y'(x_0)<-2$. It cannot yet describe the late time accelerating phase of the universe because the point is in the decelerating phase ($q>0$).\\
The third category of solutions ($P_3$ in the Table) are those for the equation $y(x)+x=1$. We call the roots of this equation $x_1$. Since both $\Omega_m$ and $\Omega_r$ are zero and $\Omega_D=1$, it will be a dark-energy-dominated solution. Both of the eigenvalues of the Jacobian matrix are negative; therefore, the fixed point is always stable. Since the deceleration parameter $q$ has a minus value for this point, it represents the accelerating phase and can successfully describe the late time accelerating stable phase of the universe.\\
The forth category of fixed points ($P_4$ in the Table) are those in which $1-\frac{1}{2}y+y'\left(x+\frac{1}{2}y\right)$ diverges and the condition $\Omega_D=x+y \neq 1$ ($P_4$ equals to $P_2$, if $\Omega_D=1$) applies. We call these points $x_2$. Given that $\Omega_m \neq 0$ in the accelerating phase ($q=-1$), the point is related to the inflation dominated era. It is a stable fixed point if $\lambda (x_2)<0$, and a saddle one if $\lambda (x_2)>0$, where $\lambda (x)$ has the following form:
\bea
\lambda (x)&=  &\frac{3}{2}\frac{(y'+2)(1-x-y)-(y'+1)(y+2x)}{1-\frac{1}{2}y+y'\left(x+\frac{1}{2}y\right)}\\
&-&\frac{3}{2}\frac{(1-x-y)(y+2x)\left(\frac{1}{2}y'(y'+1)+y''\left(x+\frac{1}{2}y\right)\right)}{\left( 1-\frac{1}{2}y+y'\left(x+\frac{1}{2}y\right)\right) ^2}.\nonumber
\eea
No situation can be imagined in which $P_4$ becomes unstable; it is, therefore not possible for the universe to have inflation exit, implying that this point cannot be accepted as a true inflation point.\\
To summarize, we have already established that there existed in the history of the universe an inflation phase, two radiation and matter-dominated eras, and a dark energy accelerating phase. As already mentioned, $P_4$ is not an unstable fixed point and cannot be, therefore, accepted as a true inflation point. As for the other three points, if we choose $f(T)$ in a way that $y'(x_0)>-2$, we will then have an unstable radiation era followed by a matter-dominated and an accelerating dark energy one. On the other hand, if $y'(x_0)<-2$, there is a saddle radiation point followed by a stable matter-dominated one and a dark energy era. In each case, one of the matter or radiation eras is not a saddle one and it will not thus be possible for the universe to remain in that situation for a long time. The case with $y'(x_0)>-2$, however, offers the advantage that the transition radiation$\rightarrow$matter$\rightarrow$dark energy will become possible. In this case matter and dark energy dominated eras will be acceptable solutions, and models of $f(T)$ will then have to be selected that have at least one point in each of the categories $P_2$ and $P_3$ presented in Table \ref{general}. In other words, each of the equations $x+y(x)=1$ and $y(x)+2x=0$ should have at least one real solution. Based on the energy conditions, we should have $0\leq\Omega_m\leq1,0\leq\Omega_r\leq1$. Hence, the preferred value of $\Omega_D$ obtained from Equations \eqref{Omegad} and \eqref{ffr1} will be $-1\leq \Omega_D\leq 1$. Since at any point in Table \ref{general}, one of the parameters $\Omega_m$ and $\Omega_r$ is zero, the preferred value of $\Omega_D$ becomes $0\leq \Omega_D\leq 1$. This yields $-1\leq x_0\leq0$, based on Equation \eqref{Omegad} and the fact that $y(x_0)+2x_0=0$. As already mentioned, the points $P_1$ and $P_2$ become dark energy dominated fixed points for $x_0=-1$. Since the deceleration parameters of these points are positive, they refer to a decelerating universe in a dark energy dominated era and correspond to no physical situation. Thus, the lower bound of $-1\leq x_0\leq0$ will be forbidden. It then follows that a viable model that contains two parts of the standard cosmological model should satisfy the following conditions:
\begin{enumerate}
  \item The equations $x+y(x)=1$ and $y(x)+2x=0$ should have at least one real solution,
  \item $-1< x_0\leq0$,
  \item $y'(x_0)>-2$.
\end{enumerate}
The study of the phase space will be exhausted by investigating the critical points at infinity, where the dynamical variables diverge. This requires the prior investigation of the valid domain of the dynamical variables, $\Omega_r$ and $x$ to see if they can possibly tend to infinity. Based on the energy conditions, the energy density of radiation and matter should be positive. This means that the allowed domain of the density parameters $\Omega_r$ and $\Omega_m$ are $0\leq\Omega_r\leq1$ and $0\leq\Omega_m\leq1$ and, consequently, one of the dynamical variables ($\Omega_r$) is always finite. Considering the allowed domain of the density parameters and using Equation \eqref{ffr1}, we obtain $-1\leq x(T)+y(T)\leq1$. This inequality determines the allowed domain of $T$. If $x(T)$ diverges in this domain, Then the study of the phase space at infinity should be performed following the Poincar\'{e} central projection method \cite{poincare}.\\
\section{Special cases}
In this section, we first consider the constraint mentioned in (\ref{exception}) before we embark on the study of some specific $f(T)$ models  with two parameters \cite{specificft}. In each case, one of the model parameters will be free while the other is determined through our analysis.  We will finally examine the conditions cited in the previous section and find the preferred values for the model parameters.
\subsection{Special case: $1+f'(T)+2Tf''(T)=  0$}
The general solution to the differential Equation (\ref{exception}) is:
\bea
f(T)=  -T+c_1\sqrt{-T}+c_2. \label{excptsol}
\eea
Based on Equation (\ref{fr2}), this solution leads to the following relation between the densities:
\bea
\rho_m=  -\frac{4}{3}\rho_r.
\eea
Except in the case $\rho_m=  \rho_r=  0$, the above equation makes no physical sense and leads to the violation of energy conditions. Plugging $\rho_m=  \rho_r=  0$ and (\ref{excptsol}) in the first Friedman Equation (\ref{fr1}) yields $c_2=  0$. This is also a trivial case and leads to no physical result.

\subsection{The Power Law Model: $f(T)=  \alpha(-T)^b$}

\begin{figure*}
\begin{center}
\begin{tabular}{cc}
\includegraphics[scale= 0.8]{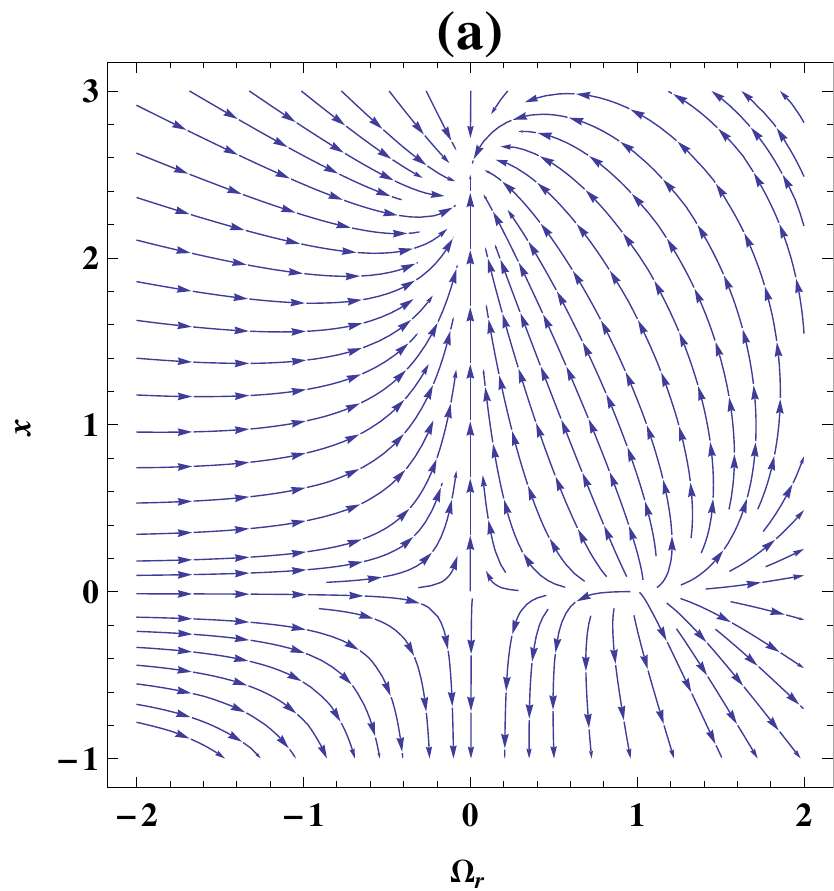}&\includegraphics[scale=0.8]{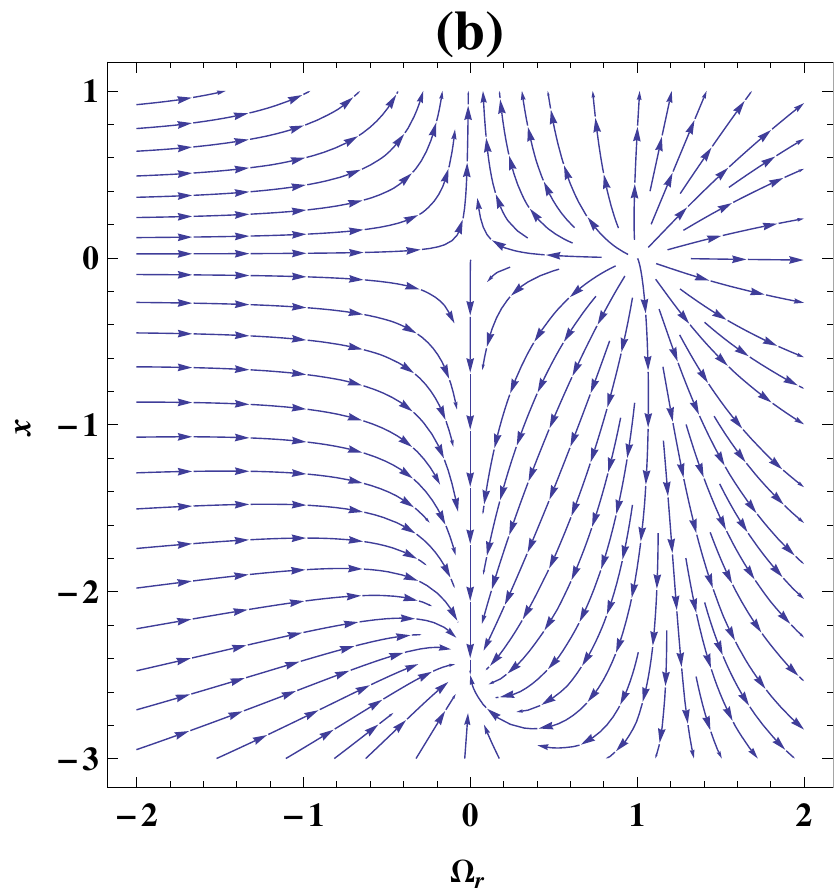}
\end{tabular}
\caption{ The behavior of the phase trajectories for the system (\ref{ode}) for $f(T)=  \alpha(-T)^b$ where $b$ has been chosen to be $0.3$ in plot \textbf{(a)} and $0.7$ in plot \textbf{(b)}. The critical points $P_1$, $P_2$, and $P_3$ are located at $(1,0)$, $(0,0)$ and $(0,\frac{1}{1-2b})$, respectively. $P_3$ is clearly a stable fixed point because the trajectories are coming out of it. $P_2$ is a saddle one because some trajectories evolve through and some of them come out of the point. Finally $P_1$ is an unstable fixed point because all of the trajectories are coming out of it. The trajectories start their evolution from $P_1$ or radiation, the only unstable fixed point in the theory. Some of the trajectories go straight forward to the stable point $P_3$ and some of them evolve first to $P_2$ or matter dominated critical point and after that to $P_3$ (dark energy era), the only attractor in the theory.}
\label{fig1}
\end{center}
\end{figure*}

A simple and useful choice for $f(T)$ is a power law function \cite{powerlaw} of the following form:
\bea
f(T)=\alpha(-T)^b, \label{powerlaw}
\eea
in which, $\alpha$ and $b$ are two model parameters. As pointed out in Refs. \cite{acceleration,powerlaw}, the model reduces to the $\Lambda$CDM model when $b=0$, and to the DGP one \cite{DGP} when $b=1/2$. However, it leads to a redefinition of the Newton constant when $b=1$. In order to be consistent with the observational data, the requirement that $|b|\ll 1$ should be observed \cite{acceleration,powerlaw,observ1,observ2}.\\
As already mentioned above, any physical function $f(T)$ should satisfy some conditions to be  a viable standard cosmological model. We will examine the power law function and see which range of $b$ will render it into a viable model for $f(T)$ gravity. The first condition to be met is that the two equations $y(x)+x=1$ and $y(x)+2x=0$ should have real solutions. For $f(T)$ in Equation (\ref{powerlaw}), we have:
\bea
&&x=  \frac{f(T)}{T}=  -\alpha(-T)^{b-1},\label{xpowerlaw}\\
&&y=  -2f'(T)=  2\alpha b(-T)^{b-1},\label{ypowerlaw}
\eea
which leads to $y(x)=-2b x$. One can check the first condition as:
\bea
&&y(x)+x=  -2b x+x=  1\Rightarrow x_1=  \frac{1}{1-2b},\\
&&y(x)+2x=  -2b x+2x=  0\Rightarrow x_0=  0.
\eea
Except for $b=\frac{1}{2}$, both of the above equations have real solutions; thus, the first condition is satisfied. The second condition states that $-1<x_0\leq 0$; since $x_0=0$, the second condition is satisfied as well. Finally, for the third condition, we have $y'(x_0)>-2$ based on \eqref{y'}:
\bea
y'(x_0)=  -2b >-2\Rightarrow b<1.
\eea
This result is consistent with earlier research \cite{acceleration,powerlaw,observ1} stating that observational data indicate that the absolute values of $b$ are far smaller than one. Thus, the conditions for matter and dark energy points are satisfied. The power law function with $b <1$ and $b\neq \frac{1}{2}$ leads to desirable results and viable cosmological predictions. Ref. \cite{observation} did not derive this constraint while the analysis provided was basically wrong. The fixed points of the power law function are plotted in Figure \ref{fig1} for two values of $b$. Three fixed points are clearly present in Figs. 1(a) and 1(b), indicating that and  the transition from a radiation era to a matter one followed by an accelerating phase is possible. Thus, the power law function successfully describes two parts of the standard cosmological model.\\
The last step is to consider the phase space to see if it is possible for the dynamical variables to tend to infinity and whether there exists any fixed point there. The allowed domain of the dynamical variables are $0\leq\Omega_r\leq1$ and $\frac{-1}{\mid 2b-1\mid}\leq x\leq\frac{1}{\mid 2b-1\mid}$ ($-1\leq x+y\leq1$). The phase space is compact except for $b=\frac{1}{2}$. For $b=\frac{1}{2}$, the dynamical variable $x$ seems to be tending to infinity and the phase space to be noncompact; this is not the case, however; rather in this case $y=-2bx=-x$ (using Eqs. \eqref{xpowerlaw} and \eqref{ypowerlaw}). The dark components in Eqs. \eqref{ffr1} and \eqref{ffr2} are , therefore, totally absent; i.e. $x+y=0$ in Eq. \eqref{ffr1} and $-\frac{1}{2}y+z=\frac{1}{2}x-(x-\frac{1}{2}x)=0$ in Eq. \eqref{ffr2}. The Friedman-like equations reduce to the typical Teleparallel theory either with no dark energy sector or no dynamical variable $x$ to tend to infinity.
\subsection{The Logarithmic Model: $f(T)=\alpha T_0\sqrt{\frac{T}{qT_0}}\ln\left(\frac{qT_0}{T}\right) $}
\begin{figure*}
\begin{center}
\begin{tabular}{cc}
\includegraphics[scale= 0.8]{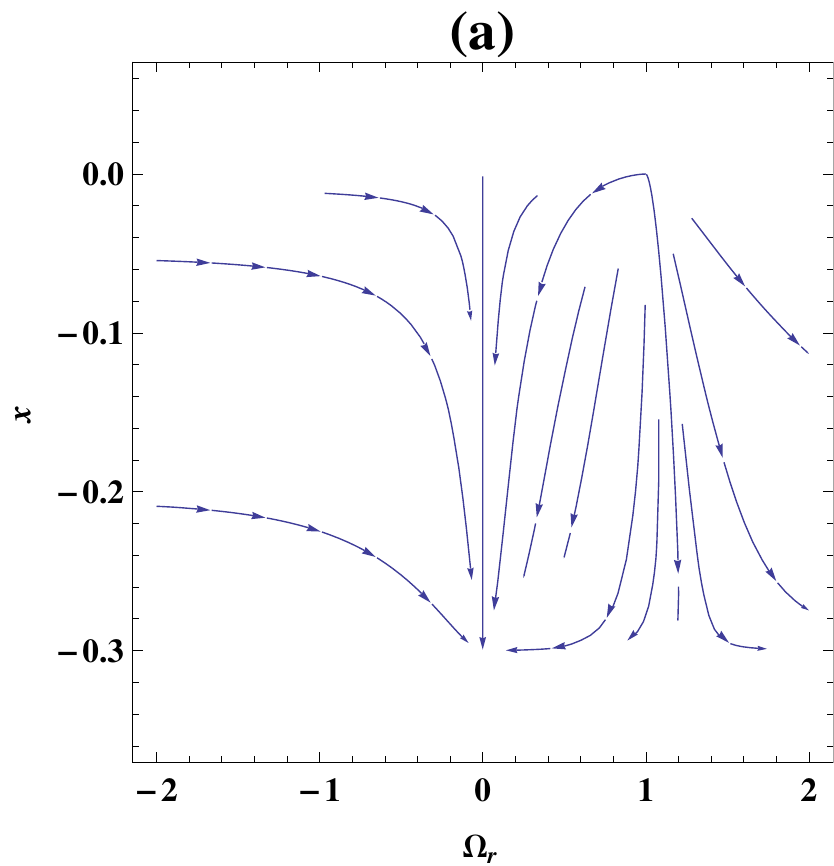}&\includegraphics[scale=0.8]{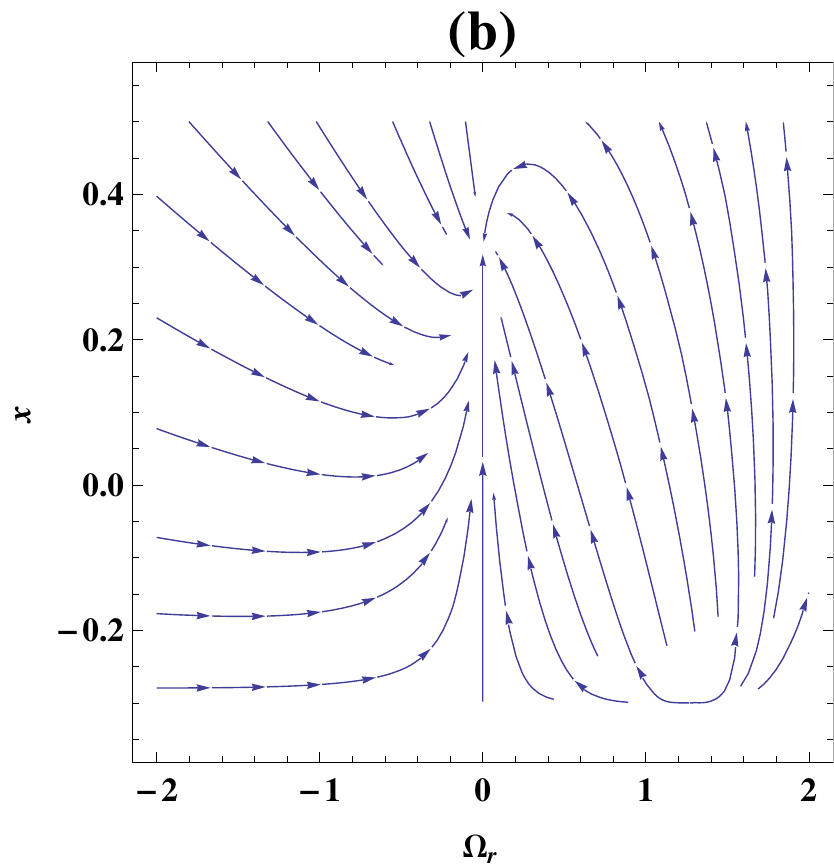}
\end{tabular}
\caption{ The behavior of phase trajectories for the system in (\ref{ode}) for $f(T)=\alpha T_0\sqrt{\frac{T}{qT_0}}\ln\left(\frac{qT_0}{T}\right) $ where $q=1$. Since the function $x(T)$ is not a one to one function (it has a minimum at $T=7.39$), it is not possible to show the whole phase space in one plot. The behavior of the phase space for $T<7.39$ and $T>7.39$ are shown in plot (a) and (b), respectively. The critical points $P_1$, $P_2$, and $P_3$ are at $(1,0)$, $(0,0)$ and $(0,0.32)$, respectively. The two unacceptable fixed points are $(1.27,-0.27)$ and $(0,-0.27)$ which exhibit different stability behaviors in the two plots.}
\label{fig2}
\end{center}
\end{figure*}

Another model is the logarithmic one \cite{logarithmic}:
\bea
f(T)=\alpha T_0\sqrt{\frac{T}{qT_0}}\ln\left(\frac{qT_0}{T}\right),\label{flog}
\eea
where, $\alpha$ and $q$ are two model parameters and $T_0$ is the present value of the torsion parameter. To study the model, we calculate $x$ and $y$ using Eqs. \eqref{x} and \eqref{y}:
\bea
x(T)&=&\frac{f(T)}{T}=\alpha\sqrt{\frac{T_0}{qT}}\ln\left(\frac{qT_0}{T}\right),\label{xlog}\\
y(T)&=&-2f'(T)=  \alpha\sqrt{\frac{T_0}{qT}}\left[2-\ln\left(\frac{qT_0}{T}\right) \right]\nonumber\\
&=&2\alpha\sqrt{\frac{T_0}{qT}}-x(T), \label{ylog}
\eea
where, $\alpha$ can be determined using \eqref{ylog}, and the present values of the cosmological parameters in  Friedman Equation \eqref{ffr1}, as follows:
\bea
\alpha=(1-\Omega_{m0}-\Omega_{r0})\frac{\sqrt{q}}{2}=0.364\sqrt{q},\label{alfalog}
\eea
where, $\Omega_{m0}=0.272$ and $\Omega_{r0}= 8.0331\times 10^{-5}$ are used \cite{omega}. According to the first condition, the following equation should have a real solution:
\bea
y(T)+2x(T)=2\alpha\sqrt{\frac{T_0}{qT}}+\alpha\sqrt{\frac{T_0}{qT}}\ln\left( \frac{qT_0}{T}\right) =0.\qquad
\eea
The above equation has two solutions at $T=-\infty$ and $T=qT_0\:e^2$. For $T=-\infty$, using \eqref{xlog} and \eqref{y'}, we obtain $x_0(T)=0$ and $y'(T)=-1$, thereby satisfying the second and third conditions ($-1<x_0\leq 0$, $y'(x_0)>-2$). For $T=qT_0\:e^2$, $x_0$ takes the following form:
\bea
x_0(T)=\alpha\sqrt{\frac{T_0}{qT}}\ln\left( \frac{qT_0}{T}\right) =-\frac{2\alpha}{qe}.\label{x0log}
\eea
As both $\alpha$ and $q$ are positive, the upper bound of $-1<x_0\leq 0$ is automatically satisfied. For the lower bound, using \eqref{x0log} and \eqref{alfalog}, we have:
\bea
x_0=-\frac{2\alpha}{eq}>-1\Rightarrow q>0.072.
\eea
For the third condition, we have:
\bea
y'(x_0)&=&\frac{dy/dT}{dx/dT}\mid_{T=qT_0e^2}=\left[ \frac{2}{2+\ln\left(\frac{qT_0}{T}\right)}-1\right] _{T=qT_0e^2}\nonumber\\&=&\pm\infty.
\eea
Since, $y'(T)$ diverges for $T=qT_0\:e^2$, the third condition is not satisfied and $x_0=-\frac{2\alpha}{qe}$ is not a true matter point. It should be noted that the place of the radiation point in this case is slightly different with from that in Table \ref{general}. Using \eqref{xlog} and \eqref{ylog}, the term $y'\left(x+\frac{1}{2}y\right)$ in the denominator of  Eq. \eqref{ode} is equal to:
\bea
&&y'(T)\left(x(T)+\frac{1}{2}y(T)\right)\nonumber\\
&=&\frac{1}{2}\left[ \frac{2}{2+\ln\left(\frac{qT_0}{T}\right)}-1\right]\alpha\sqrt{\frac{T_0}{qT}}\left( 2+\ln\left(\frac{qT_0}{T}\right)\right) \nonumber\\
&=&-\frac{1}{2}x(T).\qquad
\eea
The coordinates of the radiation point in the $(x,\Omega_r)$ plane is, therefore, $(x_0,1-x_0)$, rather than $(x_0,1+x_0)$ in Table \ref{general}. The dark energy dominated era is based on the solution of the following equation:
\bea
y(T)+x(T)=2\alpha\sqrt{\frac{T_0}{qT}}=1\Rightarrow T=  \frac{4\alpha^{2}T_0}{q}.
\eea
Since the above equation has a real solution, all the conditions are satisfied. Hence, the logarithmic form of Eq. \eqref{flog} with $q>0$ can describe successfully the matter and dark energy eras of the standard model of cosmology. The relevant phase space plot is presented in Fig. \ref{fig2} in which the five fixed points and the transition between them are clearly seen.\\
The last step is to study the infinite behavior of the dynamical variables and the related fixed points, if they ever exist. We begin with the examination of the allowed domain of the dynamical variables to see if it is possible for the dynamical variables to tend to infinity. As mentioned in the previous Section, the allowed domains of $\Omega_r$ and $x$ are, $0\leq \Omega_r \leq 1$ and $-1\leq x+y \leq 1$. The latter leads to $-1\leq 2\alpha\sqrt{\frac{T_0}{qT}}\leq 1$, based on Eq. \eqref{ylog}, and yields $T\leq\frac{4\alpha^2 T_0}{q}$. Considering Eq. \eqref{xlog}, it is obvious that, for a finite $q$, $x$ is finite in this domain and cannot tend to infinity. Hence, there are no fixed points at infinity. This concludes the study of the phase space and fixed points.
\subsection{The Hyperbolic-Tangent Model: $f(T)=\alpha (-T)^n\tanh\left( \frac{T_0}{T}\right)  $}
\begin{figure*}
\begin{center}
\includegraphics[scale= 0.8]{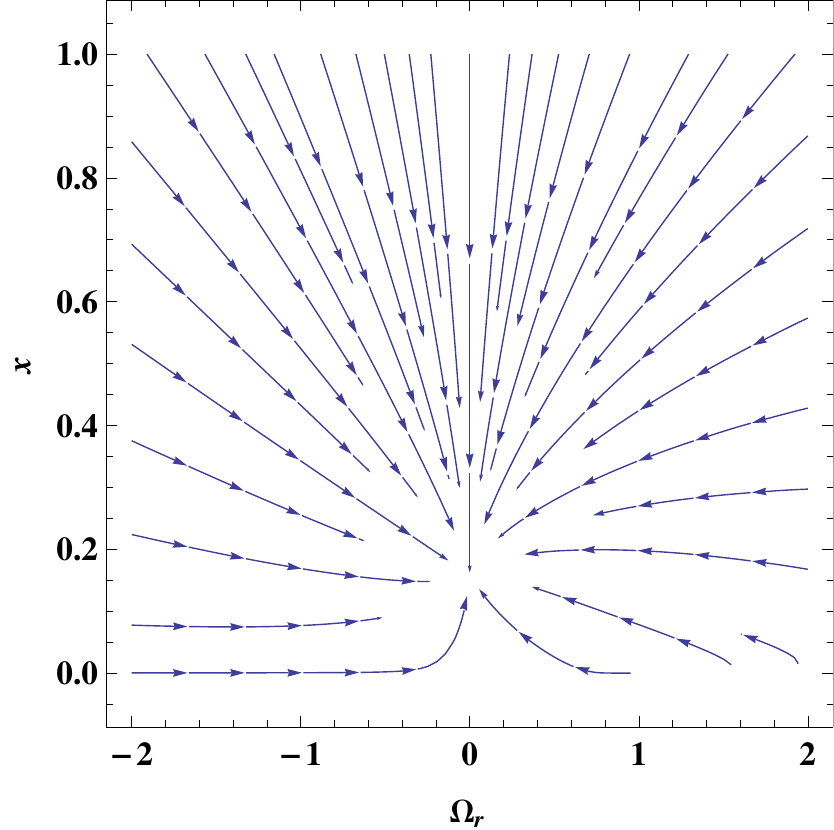}
\caption{ The behavior of the phase trajectories of the system in (\ref{ode}) for $f(T)=\alpha (-T)^n\tanh\left( \frac{T_0}{T}\right)$ where $n=-2$. The critical points $P_1$, $P_2$, and $P_3$ are located at $(1,0)$, $(0,0)$, and $(0,0.15)$, respectively.}
\label{fig3}
\end{center}
\end{figure*}
The next model to examine is the hyperbolic-tangent model \cite{hypertan} of the following form:
\bea
f(T)=\alpha (-T)^n\tanh\left( \frac{T_0}{T}\right)\label{ftan},
\eea
where, $\alpha$ and $n$ are two model parameters. Using the same calculations as in the previous subsection, we can obtain  $x$, $y$, and $\alpha$ as in the following:
\bea
&&x(T)=\frac{f(T)}{T}=-\alpha(-T)^{n-1}\tanh\left( \frac{T_0}{T}\right),\label{xtan}\\
&&y(T)=-2f'(T)\nonumber\\
&&=  2\alpha\left[ (-T)^{n-2}\cosh ^{-2}\left( \frac{T_0}{T}\right) +n(-T)^{n-1}\tanh\left( \frac{T_0}{T}\right)\right], \nonumber\\\label{ytan}\\
&&\alpha =\frac{0.728(-T_0)^{1-n}}{1.523 n + 1.601}.\label{alfatan}
\eea
To evaluate the model, we start with the matter point $x_0$, which is the solution of:
\bea
y(T)+2x(T)&=&2\alpha\Big[(-T)^{n-2}T_0\cosh ^{-2}\left( \frac{T_0}{T}\right)\nonumber\\
&+&(n-1)(-T)^{n-1}\tanh\left( \frac{T_0}{T}\right) \Big ]=0.\qquad
\eea
For $n\geq 1$ ($n< 1$), $T=0$ ($T=-\infty$) is the solution of the above equation. Since $T=0$ corresponds to $H=0$ ($T=-6H^2$), it can be a solution for the late time dark energy era; however, it cannot explain inflation, radiation, or matter eras. We, therefore, need to ignore $n\geq 1$, and keep $n< 1$ as a viable model parameter. The other conditions for the matter point are $-1<x_0\leq 0$ and $y'(x_0)>-2$, which can be checked  for $T=-\infty$ and $n< 1$ as follows:
\bea
&&x_0=\lim_{T\rightarrow -\infty} \left[ -\alpha(-T)^{n-1}\tanh\left( \frac{T_0}{T}\right)\right] =0,\nonumber\\
&&y'(x_0)=\frac{dy/dT}{dx/dT}\mid_{T=-\infty}\nonumber\\
&&=\lim_{T\rightarrow -\infty} \frac{2(1-n)T\left[ nT\sinh\left( \frac{2T_0}{T}\right)-4T_0 \right] +8T_0^2\tanh\left( \frac{T_0}{T}\right) }{T\left[(n-1)T\sinh\left( \frac{2T_0}{T}\right) -2T_0 \right] }\nonumber\\
&&=2(1-n)>-2.
\eea
Clearly the conditions for the matter point are satisfied and $T^0=-\infty$ can be accepted as a true matter solution. Now, we turn to the dark energy dominated era. The condition to be satisfied is:
\bea
&&y(T)+x(T)=\alpha\Big[2(-T)^{n-2}T_0\cosh ^{-2}\left( \frac{T_0}{T}\right)\nonumber\\
&&+(2n-1)(-T)^{n-1}\tanh\left( \frac{T_0}{T}\right) \Big] =1.
\eea
Using Eq. \eqref{alfatan} and performing numerical calculations, we find that for $n< -1$, the above equation has a real solution. The hyperbolic-tangent model with $n< -1$ is, therefore, capable of describing successfully
 the matter and dark energy eras of the standard model of cosmology. The fixed points and phase trajectories of the model are plotted in Fig. \ref{fig3}.\\
We can show that the phase space of the theory is compact and there is no infinite fixed point in the theory. This concludes the study of the phase space.
\subsection{The Linder Model: $f(T)=\alpha T_0(1-e^{-p\sqrt{\frac{T}{T_0}}})  $}
\begin{figure*}
\begin{center}
\includegraphics[scale= 0.8]{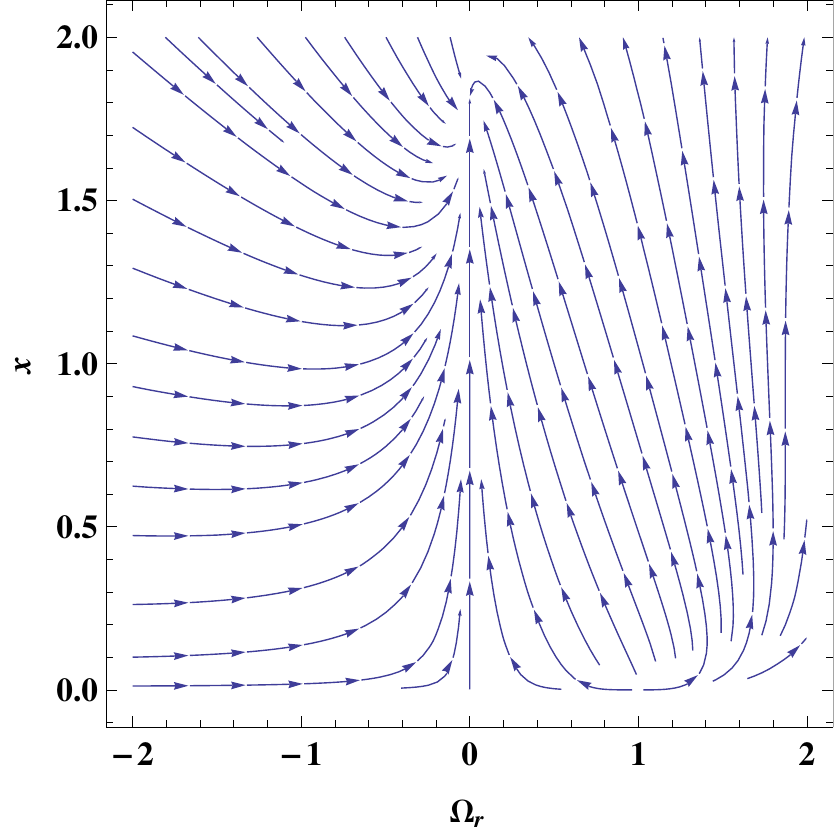}
\caption{The behavior of the phase trajectories for the Linder model, $f(T)=\alpha T_0(1-e^{-p\sqrt{\frac{T}{T_0}}})$ where, $p=2$. Critical points $P_1$, $P_2$ and $P_3$ are placed at $(1,0)$, $(0,0)$ and $(0,1.81)$ respectively.}
\label{fig4}
\end{center}
\end{figure*}

The Linder model \cite{Linder} is captured by the following function:
\bea
f(T)=\alpha T_0(1-e^{-p\sqrt{\frac{T}{T_0}}}), \label{flinder}
\eea
where, $\alpha$ and $p$ are two model parameters. $x$, $y$, and $\alpha$ may be obtained as follows:
\bea
&&x=\frac{f(T)}{T}=\alpha\frac{T_0}{T}(1-e^{-p\sqrt{\frac{T}{T_0}}}),\label{xlinder}\\
&&y=-2f'(T)=-\alpha p \sqrt{\frac{T_0}{T}}e^{-p\sqrt{\frac{T}{T_0}}}, \label{ylinder}\\
&&\alpha=\frac{1-\Omega_{m0}-\Omega_{r0}}{1-(1+p)e^{-p}}=\frac{0.728}{1-(1+p)e^{-p}}.\label{alfalinder}
\eea
Acording to the first condition, the following equation should have a real solution:
\bea
&&y(T)+2x(T)=2\alpha\frac{T_0}{T}(1-e^{-p\sqrt{\frac{T}{T_0}}})-\alpha p \sqrt{\frac{T_0}{T}}e^{-p\sqrt{\frac{T}{T_0}}} \nonumber\\
&&=0.\label{x0linder}
\eea
Numerical calculations show that the above equation has a root at $T=\frac{2.54T_0}{p^2}$ for $p<0$. The other conditions for the matter point are $-1<x_0\leq 0$ and $y'(x_0)>-2$. Replacing $T=\frac{2.54T_0}{p^2}$ in (\ref{xlinder}) yields $x_0=\frac{0.287(1-e^{-1.59\frac{p}{\mid p\mid}})p^2}{1-e^{-p}(1+p)}$, which is in the desired regime for $p<-1.162$. For a negative $p$, however, $y'(x_0)$ is always less than $-2$. Hence, the conditions for the matter point are not satisfied for $p<0$. For $p>0$, $T=-\infty$ is the solution of \eqref{x0linder}. Since $x(-\infty)=y'(-\infty)=0$, both conditions for the matter point are satisfied for $p>0$. We turn to the dark energy point $x_1$, which is the solution of the following equation:
\bea
&&y(T)+x(T)=\alpha p \sqrt{\frac{T_0}{T}}e^{-p\sqrt{\frac{T}{T_0}}}+\alpha \frac{T_0}{T}\left( 1-e^{-p\sqrt{\frac{T}{T_0}}}\right) \nonumber\\&&=1.
\eea
Numerical calculations show that the above equation has a real solution except for $0\leq p<0.5$. It follows that the Linder model with $p\geq 0.5$ is capable of describing successfully the matter and dark energy dominated eras of the universe. Phase space trajectories of the Linder model for $p=2$ are plotted in Fig. \ref{fig4}, in which radiation, matter, and dark energy fixed points and the transition between them can be identified in Fig. \ref{fig4}.\\
As in the previous models, may be claimed that the phase space is compact and there exist no infinite fixed points.
\subsection{The Exponential Model: $f(T)=\alpha T_0(1-e^{-p\frac{T}{T_0}})  $}
\begin{figure*}
\begin{center}
\begin{tabular}{cc}
\includegraphics[scale= 0.83]{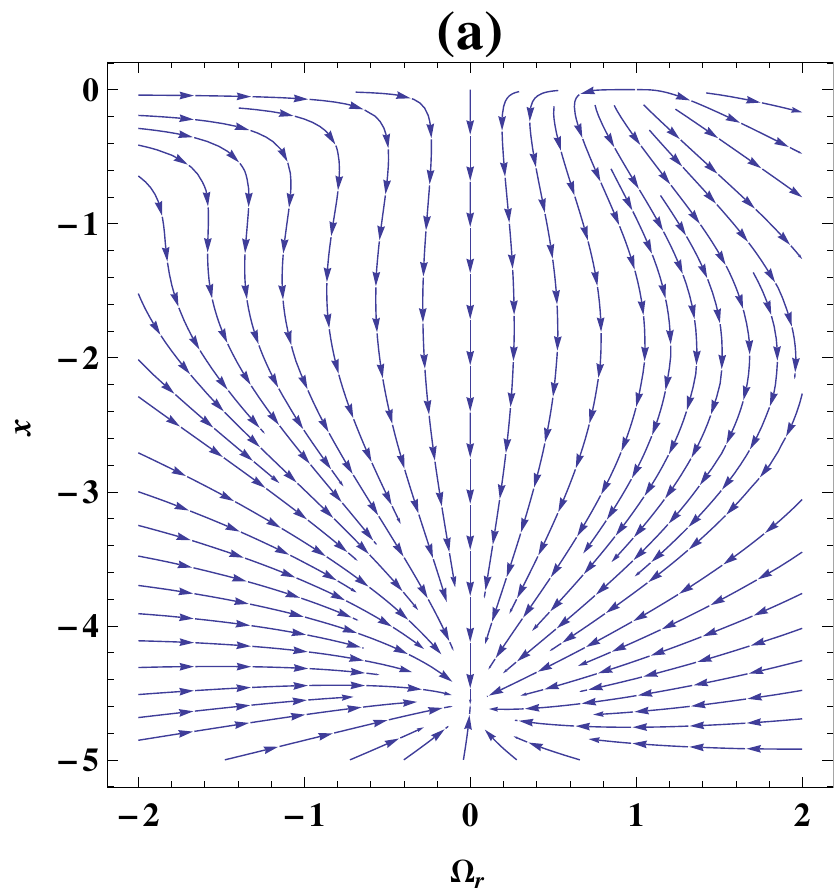}&\includegraphics[scale=0.8]{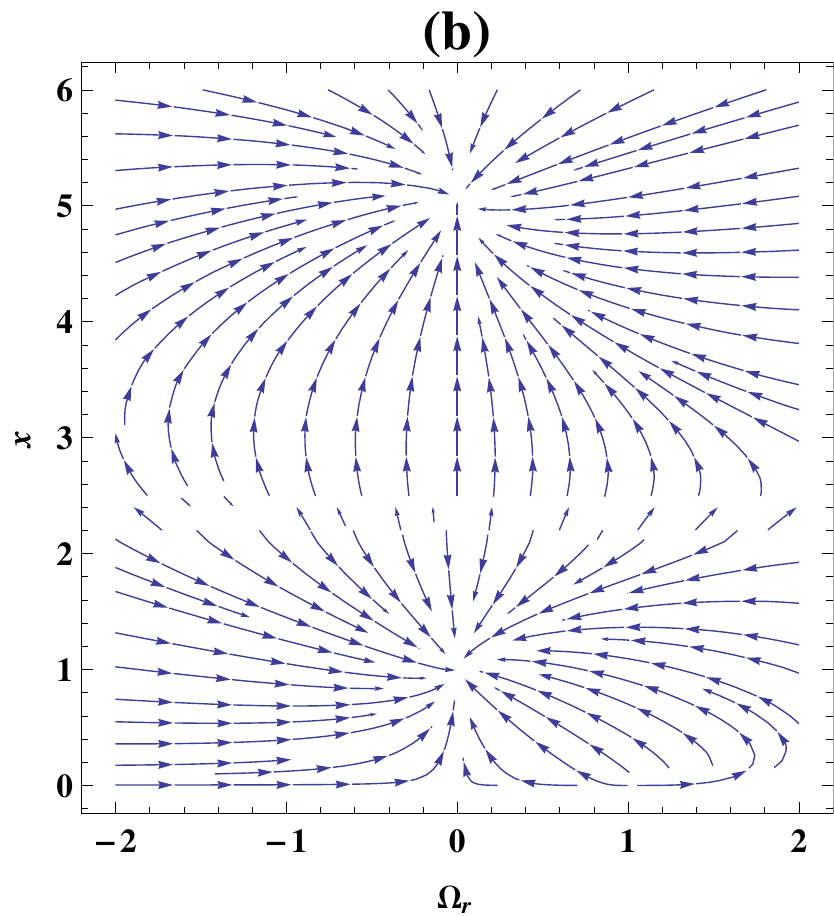}
\end{tabular}
\caption{The behavior of the phase trajectories for the exponential model, $f(T)=\alpha T_0(1-e^{-p\frac{T}{T_0}})$ where, $p=1$ in plot (a) and $p=1.5$ in plot (b). The critical points $P_1$ and $P_2$ are located at $(1,0)$, and $(0,0)$, respectively. For $0.203\leq p\leq 1.256$, there is one $P_3$ at $(0,-4.59)$ in plot (a). For $1.256\leq p\leq 1.977$, there are two $P_3$ solutions located at at $(0,1)$ and $(0,5.04)$ in plot (b). The late time attractor of the universe is $(0,1)$, to which the transition of the matter point.}
\label{fig5}
\end{center}
\end{figure*}

Finally, we investigate the exponential model \cite{exponential} defined as follows:
\bea
f(T)=\alpha T_0(1-e^{-p\frac{T}{T_0}}) \label{fexp},
\eea
where, $\alpha$ and $p$ are two model parameters. $x$, $y$, and $\alpha$are expressed as follows:
\bea
&&x=\frac{f(T)}{T}=\alpha\frac{T_0}{T}(1-e^{-p\frac{T}{T_0}}),\label{xexp}\\
&&y=-2f'(T)=-2\alpha p e^{-p\frac{T}{T_0}}, \label{yexp}\\
&&\alpha =\frac{1-\Omega_{m0}-\Omega_{r0}}{1-(1+2p)e^{-p}}=\frac{0.728}{1-(1+2p)e^{-p}}.\label{alfaexp}
\eea
The matter point, $x_0$ is the solution of the following equation:
\bea
&&y(T)+2x(T)=-2\alpha p e^{-p\frac{T}{T_0}}+2\alpha\frac{T_0}{T}(1-e^{-p\frac{T}{T_0}})\nonumber\\&&=0.
\eea
The above equation, has two solutions at $T=0$ and $T=-\infty$ for $p>0$. $T=0$ is not an acceptable solution for the matter point while for $T=-\infty$, $x_0=y'(x_0)=0$, and the conditions for the matter point are satisfied as well. The condition for the dark energy dominated era is that the solution for the following equation should exist:
\bea
y(T)+x(T)=-2\alpha p e^{-p\frac{T}{T_0}}+\alpha\frac{T_0}{T}(1-e^{-p\frac{T}{T_0}}) =1.\qquad
\eea
The function $x(T)+y(T)$ has a complicated behavior. For $0\leq p<0.203$, the above equation has a positive solution ($T>0$); however, the solution is not acceptable since $T=-6H^2$. The function $x(T)+y(T)$ has a minimum and a maximum value for $ 0.203\geq p<1.256 $ and $p>1.256 $, respectively. For $p=1.256$, $\alpha$ becomes infinite. For $1.256\leq p\leq 1.977$, the maximum value of the $x(T)+y(T)$ is grater than $1$; hence, there are two solutions for $x(T)+y(T)=1$. This leads to the question which solution is the late time attractor of the universe. As the valid range of $T$ breaks up into two parts in the case, the true domain is the one that contains $x_0$; i.e., the lower value of $T$ (for $x_0$, $T=-\infty$). For $1.977<p<6.110$, the maximum value of  $x(T)+y(T)$ is less than $1$ and there is, thus, no solution at all. Hence, the valid domains of $p$, in which the theory explains both matter and dark energy dominated eras, are $0.203\leq p\leq 1.977$, $p\geq 6.110$ and $p\neq 1.256$. In Fig. \ref{fig5}, the behavior of the phase trajectories and fixed points of the exponential model are plotted for two values of $p$.\\
It is observed that the infinite value for $x$ is forbidden and the phase space is compact in this model, too. We conclude that the study of the phase space and the fixed points of the theory is complete.\\
Table \ref{special} summarizes the general features and the related domains of the independent model parameters of the five specific two-parameter models investigated in this section. It is straightforward to check other functional forms of the $f(T)$ theory using similar calculations. Once the $f(T)$ function is chosen, it will be possible to calculate the variables $x$ and $y$ along the lines outlined above and calculate $x_0$ and $x_1$ either analytically or numerically. The existence of a real $x_0$ is a necessary condition for describing the matter dominated era. The two other conditions, which are $-1<x_0\leq 0$ and $y'(x_0)>-2$, should be satisfied as well. However, the existence of a real $x_1$ suffices for describing the dark energy dominated era.
\begin{table*}
\begin{center}
\caption{\bf General features and the related domains of the independent model parameters of specific $f(T)$ functions.}
\begin{tabular}{|c|c|c|}
\hline

  {\bf The model} & {\bf $f(T)$ function} &  {\bf Allowed range of the independent parameter}\\\hline\hline
Power law & $\alpha(-T)^b$ & $b<1,b \neq \frac{1}{2}$ \\\hline

Logarithmic & $\alpha T_0\sqrt{\frac{T}{qT_0}}\ln\left(\frac{qT_0}{T}\right) $ & $q>0$ \\\hline

Hyperbolic-tangent & $\alpha (-T)^n\tanh\left( \frac{T_0}{T}\right)$ & $n<-1$ \\\hline

Linder & $\alpha T_0(1-e^{-p\sqrt{\frac{T}{T_0}}})$ & $p\geq 0.5$   \\\hline

Exponential & $\alpha T_0(1-e^{-p\frac{T}{T_0}})$ & $0.203\leq p\leq 1.977$, $p\geq 6.110$, $p\neq 1.256$  \\\hline

\end{tabular}
\label{special}
\end{center}
\end{table*}
\section{Concluding remarks}
The cosmological behavior of the $f(T)$ gravity theory was investigated for the perfect dust matter and radiation in the flat FRW universe using the dynamical systems method. It was found that the $f(T)$ theory was able to describe two periods of the standard cosmological model, namely, matter and accelerating dark energy dominated eras, if the following three conditions are satisfied. The first condition concerns the existence of at least one fixed point for any one of the periods mentioned. The second accounts for the saddle behavior of the matter fixed point. The third is related to the energy conditions. Since $\Omega_m$ and $\Omega_r$ are proportional to $\rho_m$ and $\rho_r$, negative values of $\Omega_m$ and $\Omega_r$ are forbidden and the theory should predict $0\leq \Omega_m\leq 1,0\leq\Omega_r\leq 1$. Based on these conditions, the theory describes successfully two periods of standard cosmological model, namely matter and dark energy dominated eras. The constraint $1+f'(T)+2Tf''(T)=0$ was studied separately. Five special cases of the $f(T)$ function (namely the power law, logarithmic, hyperbolic-tangent, Linder, and exponential models) were also investigated. All the models were found to satisfy the conditions for certain values of the model parameters, which revealed their capability to describe the matter and dark energy eras of the standard model of cosmology. Our method may also be used for other generalized theories of gravity.


\begin{thebibliography}{99}
\bibitem{astro}
A. G. Riess et al., {Astron. J. } {\bf 116} (1998) 1009;\\
S. Perlmutter et al., {Astrophys. J. } {\bf 517} (1999) 565;\\
P.deBernardis et al., {Nature} {\bf 404} (2000)955;\\
S. Perlmutter etal., {Astrophys. J. } {\bf 598} (2003) 102.
\bibitem{weits}
Weitzenb\"{o}ck R., Invarianten Theorie, (Nordhoff, Groningen,1923).
\bibitem{fr}
S. Nojiri and S. D. Odintsov, {Phys. Rept. } {\bf 505} (2011) 59-144;\\
S. M. Carroll, V. Duvvuri, M. Trodden and M. S. Turner, {Phys. Rev. D } {\bf 70} (2001) 043528;\\
S. Nojiri and S. D. Odintsov, {Phys. Lett. B } {\bf 657} (2007) 238; {Phys. Rev. D } {\bf 77} (2008) 026007;\\
G. Cognola,  E. Elizalde, S. Nojiri, S. D. Odintsov, L. Sebastiani and S. Zerbini, {Phys. Rev. D } {\bf 77} (2008) 046009; {\bf 83} (2011) 086006;\\
S.Capozziello, V.F.Cardone and A.Troisi, {J. Cosmol. Astropart. Phys. } {\bf 08} (2006) 001; {Mon. Not. R. Astron. Soc. } {\bf 375} (2007) 1423;\\
A. Borowiec, W. Godlowski and M. Szydlowski, {Int. J. Geom. Methods Mod. Phys. } {\bf 4} (2007) 183;\\
C. F. Martins and P. Salucci, {Mon. Not. R. Astron. Soc. } {\bf 381} (2007) 1103;\\
C. G. Boehmer, T. Harko and F. S. N. Lobo, {Astropart. Phys. } {\bf 29} (2008) 386;\\
C. G. Boehmer, T. Harko and F. S. N. Lobo, {Astropart. Phys. } {\bf 03} (2008) 024.
\bibitem{high}
S. Nojiri, S. D. Odintsov and M. Sasaki, Phys. Rev. D {\bf 71} (2005) 123509 [hep-th/0504052],\\
S. Nojiri and S. D. Odintsov, J. Phys. Conf. Ser. {\bf 66}, 012005 (2007).
\bibitem{teleparallel}
K. Hayashi and T. Shirafuji, Phys. Rev. D {\bf19}, 3524 (1979); Addendum-ibid. {\bf24}, 3312 (1982).
\bibitem{TEGR}
A. Einstein 1928, Sitz. Preuss. Akad. Wiss. p. 217; ibid p. 224;\\
A. Unzicker and T. Case, arXiv:physics/0503046.
\bibitem{gauge}
V. C. de Andrade, L. C. T. Guillen, and J. G. Pereira, Phys. Rev. Lett. {\bf84}, 4533 (2000).
\bibitem{ft}
G. R. Bengochea and R. Ferraro, Phys. Rev. D {\bf79}, 124019 (2009);\\
E. V. Linder, Phys. Rev. D {\bf81}, 127301 (2010), arXiv:1005.3039 [astro-ph.CO].
\bibitem{revft}
Y. F. Cai, S. Capozziello, M. D. Laurentis, E. N. Saridakis, Rept.Prog.Phys. 79 (2016) no.10, 106901, arXiv:1511.07586 [gr-qc].
\bibitem{acceleration}
G. R. Bengochea and R. Ferraro, Phys. Rev. D {\bf79}, 124019 (2009).
\bibitem{inflation}
R. Ferraro and F. Fiorini, Phys. Rev. D {\bf 75}, 084031 (2007), arXiv:gr-qc/0610067;\\
R. Ferraro and F. Fiorini, Phys. Rev. D {\bf 78}, 124019 (2008), arXiv:0812.1981v1 [gr-qc].
\bibitem{cosmology}
R. Myrzakulov, Eur. Phys. J. C {\bf71}, 1752, (2011), arXiv:1006.1120 [gr-qc];\\
J. B. Dent, S. Dutta and E. N. Saridakis, JCAP {\bf1101}, 009, (2011), arXiv:1010.2215 [astro-ph.CO].
\bibitem{scalar}
K. K. Yerzhanov, S. R. Myrzakul, I. I. Kulnazarov and R. Myrzakulov, arXiv:1006.3879 [gr-qc].
\bibitem{observ1}
P. Wu and H. Yu, Phys. Lett. B {\bf 693}, 415-420, (2010), arXiv:1006.0674.
\bibitem{Neother}
H. Wei, X. J. Guo and L. F.  Wang, Phys. Lett. B {\bf707}, 298-304, (2012), arXiv:1112.2270 [gr-qc];\\
K. Atazadeh and F. Darabi1, Eur. Phys. J. C {\bf72}, 2016, (2012), arXiv:1112.2824[physics.gen-ph].
\bibitem{wormhole}
M. Jamil, D. Momeni and R. Myrzakulov, Eur. Phys. J. C  {\bf73}, 2267, (2013), arXiv:1212.6017[gr-qc].
\bibitem{phantomcross}
P.Wu and H.W. Yu, Eur. Phys. J. C {\bf71}, 1552, (2011), arXiv:1008.3669 [gr-qc];\\
K. Karami and A. Abdolmaleki, arXiv:1009.2459 [gr-qc].
\bibitem{fix}
C. G. B\"{o}hmer and N. Chan, Res. Astron. Astrophys. {\bf13}, 757, (2013), arXiv:1409.5585[gr-qc].
\bibitem{fix ft}
Y. Zhang, H. Li, Y. Gong and Z. H. Zhu, JCAP, {\bf07}, 015, (2011), arXiv:1103.0719[astro-ph.CO];\\
C. Xu, E. N. Saridakisb and G. Leon, arXiv:1202.3781[gr-qc];\\
M. Jamil, D. Momeni and R. Myrzakulov, Eur. Phys. J. C, {\bf72}, 1959, (2012), arXiv:1202.4926[physics.gen-ph];\\
M. Jamil, K. Yesmakhanova, D. Momeni and R. Myrzakulov, Cent. Eur. J. Phys. {\bf 10}, 1065-1071, (2012), arXiv:1207.2735[gr-qc];\\
S. Kr. Biswas and S. Chakraborty, arXiv:1504.02431[gr-qc];\\
S. Carloni, F. S. N. Lobo, G. Otalora, E. N. Saridakis, Phys. Rev. D 93 (2016) 024034, arXiv:1512.06996 [gr-qc];\\
M. Skugoreva, E. Saridakis, A. Toporensky, Phys. Rev. D 91 (2015) 044023, arXive:1412.1502 [gr-qc];\\
G. Kofinas, G. Leon, E. N. Saridakis, Class. Quant. Grav. 31 (2014) 175011, arXiv:1404.7100 [gr-qc].
\bibitem{observation}
P. Wu and H. Yu,  Phys. Lett. B {\bf 692}, 176 (2010), arXiv:1007.2348 [astro-ph.CO].
\bibitem{poincare}
S. Lynch, Dynamical Systems with Applications using Mathematica, Birkhauser, Boston (2007).
\bibitem{specificft}
S. Nesseris, S. Basilakos, E. N. Saridakis, L. Perivolaropoulos, Phys. Rev. D {\bf 88}, 103010 (2013), arXiv:1308.6142.
\bibitem{powerlaw}
G. R. Bengochea and R. Ferraro, Phys. Rev. D, {\bf 79}, 124019, (2009).\\
E. V. Linder, Phys. Rev. D {\bf 81}, 127301 (2010).
\bibitem{DGP}
G. Dvali, G. Gabadadze, M. Porrati, Phys. Lett. B {\bf 485}, 208 (2000).
\bibitem{observ2}
R. C. Nunes, S. Pan, E. N. Saridakis, JCAP 1608 (2016) no. 08, 011, arXiv:1606.04359 [gr-qc];\\
R. C. Nunes, A. Bonilla, S. Pan, and E. N. Saridakis, arXiv:1608.01960 [gr-qc].
\bibitem{logarithmic}
K. Bamba, G. C. Qiang, L. C. Chi, L. L. Wei, JCAP 1101, 021 (2011).
\bibitem{omega}
G. Hinshaw et al. [WMAP Collaboration], Astrophys. J. Suppl. 208, 19 (2013), arXiv:1212.5226 [astro-ph.CO].
\bibitem{hypertan}
P. Wu and H. W. Yu, Eur. Phys. J. C {\bf 71}, 1552 (2011).
\bibitem{Linder}
E. V. Linder, Phys. Rev. D {\bf 81}, 127301 (2010); Erratum, Phys. Rev. D, {\bf 82}, 109902.
\bibitem{exponential}
E. V. Linder, Phys. Rev. D, {\bf 80}, 123528, (2009).
\end{thebibliography}
\end{document}